\newcounter{myctr}
\def\myitem{\refstepcounter{myctr}\bibfont\noindent\ifnum\themyctr>9\else\phantom{0}\fi\hangindent17pt\themyctr.\enskip}

\documentclass{ws-ijqi}
\usepackage{hyperref}
\usepackage[super,sort,compress]{cite}
\usepackage{hyperref}
\begin{document}

\catchline{}{}{}{}{}

\title{GENERATION AND TELEPORTATION OF THREE AND FOUR PARTICLE W STATE
}

\author{Seyed Amir hossein Mehrinezhad Chobari}
\address{Department of Electrical and Computer Engineering, University of Tehran, Tehran, Iran}
\author{Hossein Aghababa }
\address{Department of Engineering, Loyola University Maryland, Maryland, USA\\
Founder of Quantum Computation and Communication Laboratory (QCCL), University of Tehran,Tehran, Iran}
\author{Mohammadreza Kolahdouz}
\address{Department of Electrical and Computer Engineering, University of Tehran, Tehran, Iran}

\maketitle

\begin{abstract}
In this paper, we introduced circuits for three- and four-particle quantum systems to generate W states with any arbitrary coefficients and phases. Subsequently, each qubit was transmitted separately through a four-qubit entangled channel. Before transmission, the sender performed pre-processing on their qubits to minimize the resources required for transmission. Additionally, the receiver applied post-processing using the ancilla qubit(s) to recover the final states.
To further improve efficiency, it is preferable to implement the protocol in a bidirectional manner, as this allows the unknown qubits initially held by the users to be utilized ancilla qubit(s). Finally, we compared our protocol with similar works and validated the correctness of the protocol by simulating it using Qiskit, a tool provided by IBM.

\end{abstract}



\markboth{\textit{Generation and Teleportation of three and four particle W state}}
{\textit{Generation and Teleportation of three and four particle W state}}
\section{Introduction}	
With advancements in quantum technologies, there is a growing need to evolve quantum communication protocols. Recently,in the field of quantum broadcasting \citen{Mafi2023broadcast,Mafi2024broadcast}, quantum secret sharing \cite{Khorrampanah2022} And one of the most important of them is quantum teleportation.\\
Quantum teleportation is a protocol for transferring the quantum state of one or multiple separate or entangled qubits without physically transmitting them. It was first proposed theoretically in 1993 by Bennett and his colleagues  \cite{bennett1993teleporting} . The first experimental implementation was carried out in 1997 by Anton Zeilinger using optical devices \cite{Bouwmeester1997} . Following this practical realization, extensive research has been conducted to advance quantum protocols further.\\
A bidirectional protocol in which two qubits are transmitted simultaneously between two users\cite{sadeghi2017}, a bidirectional protocol that allows the transmission of an arbitrary number of qubits between two users \cite{Zadeh2017} and also, the channel is noisy \cite{Zadeh2019},Performance Analysis of Hardware-Efficient Algorithms in Noisy Intermediate-Scale Quantum Devices\cite{Ahmadkhaniha2023mafi} ,an arbitrary number of qubits in the GHZ state  \cite{Kazemikhah2021}, a controlled asymmetric protocol where two qubits are sent from one side and three qubits are sent from the other side \cite{Kazemikhah2022} ,bidirectional quantum teleportation of an arbitrary number of qubits over a noisy quantum channel\cite{Mafi2022}, an asymmetric three-party protocol \cite{Mahjoory2023} ,Quantum teleportation with distributed gates \cite{Ahmadkhaniha2023}, a controlled protocol with the ability to change the receiver \cite{Sadeghizadeh2023} and also, a bidirectional controlled protocol for an arbitrary number of qubits transmitted through a multi-hop network.\cite{Mafi2024} \\
Quantum teleportation is performed through an entangled channel. W states have been used as a channel in numerous studies. The standard W quantum state for an arbitrary number of $n$ qubits is represented as $|W_n \rangle = \frac{1}{\sqrt{n}} \left( |100\ldots 0 \rangle + |010\ldots 0 \rangle + \cdots + |000\ldots 1 \rangle \right)$
Articles \citen{Joo20031,Shi2002,Cao2007}  have proposed schematics for teleporting an unknown qubit through a W state,examines the effect of noise on teleportation when using GHZ or W states \cite{Jung2008} ,  multi-hop teleportation protocols based on W states and EPR pairs. \citen{Zhan2016,Zhang2018} a protocol for teleporting an arbitrary and unknown two-qubit state from a sender to one of two receivers using W state and GHZ state. \cite{Dai2004} ,transmits two qubits through channels based on pseudo-GHZ and W states. \cite{Nusur2021} \\
Numerous articles have explored the generation of W states on various platforms, including superconducting platforms \citen{Zhang2023,Neeley2010,Hofheinz2009}, optical platforms \citen{Gräfe2014,Zou2002,Mikami2004}, and trapped ions \citen{Cole2021,Sharma2008,Amniat2012}.
Also, the theoretical construction of the GHZ state\cite{Mafi2024ghz} And the detector of entangled state using machine learning \cite{Kookani2024}  has been introduced.
In the second section of this paper, we generated the W state for three-particle systems, and in the third section, the W state for four-particle systems using quantum circuits,In the fourth section, we addressed the teleportation of the W state for three-particle systems, and in the fifth section, we focused on the teleportation of the W state for four-particle systems,to increase efficiency in these protocols, we employed data processing techniques. This processing is particularly useful for multi-hop protocols, as it minimizes the quantum and classical resources required at the repeater stage,in section six, we demonstrated that the efficiency further increases if the protocol is bidirectional. This is because qubits that were initially in the W state and later transformed to the $|0\rangle $ state after processing can be used as ancilla qubits ,In the final section, we compare the efficiency of the protocols introduced in this paper with similar works.

\section{Generation of a Three-Particle W State  }
\begin{figure}[h]
\centerline{\includegraphics[width=0.80\textwidth]{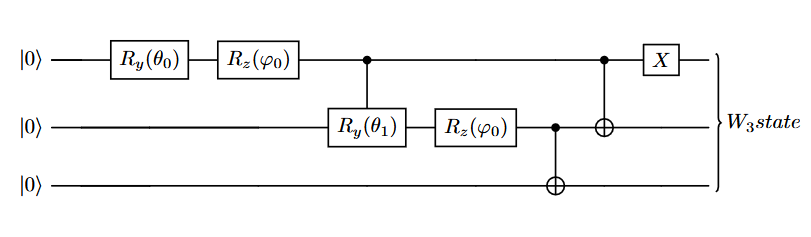}}
\vspace*{8pt}
\caption{ Circuit for generating the three-particle W state}
\label{fig:1}
\end{figure}

Initially, the first qubit is initialized using the Ry($\theta_0$) and Rz($\varphi_0$) gates to bring it to the  $|\psi\rangle = \left( \cos\left(\frac{\theta_0}{2}\right) |0\rangle - e^{i\varphi_0} \sin\left(\frac{\theta_0}{2}\right) |1\rangle \right) |00\rangle$ state,   Next, a controlled-Ry($\theta_1$) gate is applied between the first( as the controller) and second(as the target) qubits to create a linear superposition of three states,  a Rz($\varphi_1$)  gate is then applied to introduce a phase on the second qubit, allowing for arbitrary phases to be applied to all states,$|\psi\rangle = \cos\left(\frac{\theta_0}{2}\right) |00\rangle - e^{i\varphi_0} \sin\left(\frac{\theta_0}{2}\right) \cos\left(\frac{\theta_1}{2}\right) |10\rangle + e^{i(\varphi_0 + \varphi_1)} \sin\left(\frac{\theta_0}{2}\right) \sin\left(\frac{\theta_1}{2}\right) |11\rangle
$then, the CNOT (second qubit, third qubit), CNOT (first qubit, second qubit), and X (first qubit) gates are applied as shown in Fig.\ref{fig:1} to obtain the final state :

\begin{equation}
|\psi\rangle = \alpha_0 |100\rangle + \alpha_1 |010\rangle + \alpha_2 |001\rangle  
 \label{1}
\end{equation}
where the Eq. \eqref{1}$ ;\  \alpha_0=\cos\left(\frac{\theta_0}{2}\right)$,$\alpha_1 = -e^{i\varphi_0} \sin\left(\frac{\theta_0}{2}\right) \cos\left(\frac{\theta_1}{2}\right)$,$\alpha_2 = e^{i(\varphi_0 + \varphi_1)} \sin\left(\frac{\theta_0}{2}\right) \sin\left(\frac{\theta_1}{2}\right)$Additionally, the coefficients are normalized$|\alpha_0|^2 + |\alpha_1|^2 + |\alpha_2|^2 =1$

\section{Generation of a Four-Particle W State}

\begin{figure}[t]
\centerline{\includegraphics[width=0.80\textwidth]{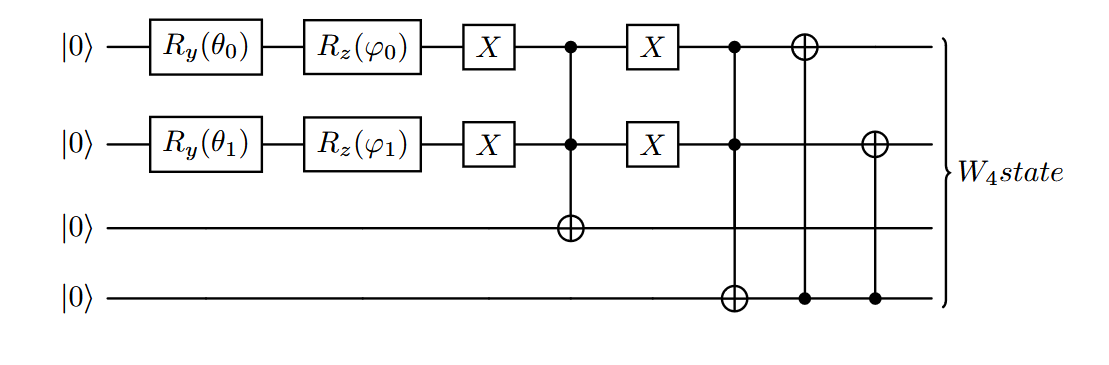}}
\vspace*{8pt}
\caption{ Circuit  for generating the four-particle W state}
\label{fig:2}
\end{figure}

The state of the system after applying the Ry($\theta_0$) , Rz($\varphi_0$) , Ry($\theta_1$) and Rz($\varphi_1$) gates to initialize the first and second qubits will be as follows: 
$|\psi\rangle =\left(\cos\left(\frac{\theta_0}{2}\right) |0\rangle -e^{i\varphi_0}\sin\left(\frac{\theta_0}{2}\right) |1\rangle\right)\left(\cos\left(\frac{\theta_1}{2}\right)|0\rangle-e^{i\varphi_1}\sin\left(\frac{\theta_1}{2}\right) |1\rangle \right)|00\rangle$Subsequently, by applying the gates shown in Fig.\ref{fig:2} , we arrive at the Eq. \ref{2}

\begin{equation}
|\psi\rangle = \beta_0 |0010\rangle + \beta_1 |0100\rangle + \beta_2 |1000\rangle + \beta_3 |0001\rangle  
 \label{2}
\end{equation}
where in the relation,$\beta_0 = \cos\left(\frac{\theta_0}{2}\right) \cos\left(\frac{\theta_1}{2}\right)$, $\beta_1 = -e^{i\varphi_1} \cos\left(\frac{\theta_0}{2}\right) \sin\left(\frac{\theta_1}{2}\right)$ , $
\beta_2 = -e^{i\varphi_0} \sin\left(\frac{\theta_0}{2}\right) \cos\left(\frac{\theta_1}{2}\right)$ , $\beta_3 = e^{i(\varphi_0 + \varphi_1)} \sin\left(\frac{\theta_0}{2}\right) \sin\left(\frac{\theta_1}{2}\right)$And also, the coefficients are normalized $|\beta_0|^2 + |\beta_1|^2 + |\beta_2|^2 + |\beta_3|^2 = 1$

\section{Teleportation   of a Three-Particle W State }
Suppose Alice has the quantum state $|\phi\rangle_{\text{A}_0 \text{A}_1 \text{A}_3}= \alpha_0 |100\rangle + \alpha_1 |010\rangle + \alpha_2 |001\rangle$ and wants to transmit this quantum state to Bob, To achieve this, Alice first performs a pre-processing with the X(first qubit), CNOT (first qubit, second qubit), CNOT (second qubit, third qubit)   gates on her qubits to assign the unknown coefficients to her first and second qubits, sending only those two. The third qubit will be set to $|0\rangle$  and will be disregarded.    Thus, Alice's quantum state becomes as follows:$ |\phi'\rangle_{\text{A}_0 \text{A}_1}= (\alpha_0 |00\rangle + \alpha_1 |10\rangle + \alpha_2 |11\rangle)$ \\
For the quantum channel, each user shares two qubits($a_0$,$a_1$ by Alice and $b_0$,$b_1$ by Bob), preparing the qubits in the quantum $
|\omega\rangle_{\text{a}_0 \text{b}_0 \text{a}_1 \text{b}_1} = 
\frac{1}{2} \left( |0000\rangle + |0101\rangle + |1010\rangle + |1111\rangle \right)$\\
Thus, the entire system$|\phi'\rangle \otimes |\omega\rangle$  can be expressed as Eq. \eqref{3}. 

\begin{equation}
\begin{aligned}
|\psi\rangle_{\text{A}_0 \text{A}_1 \text{a}_0 \text{a}_1 \text{b}_0 \text{b}_1} = & \frac{1}{2} \left( 
\alpha_0 |000000\rangle + \alpha_0 |000101\rangle + \alpha_0 |001010\rangle + \alpha_0 |001111\rangle \right. \\
& \left. + \alpha_1 |100000\rangle + \alpha_1 |100101\rangle + \alpha_1 |101010\rangle + \alpha_1 |101111\rangle \right. \\
& \left. + \alpha_2 |110000\rangle + \alpha_2 |110101\rangle + \alpha_2 |111010\rangle + \alpha_2 |111111\rangle 
\right)
 \label{3}
\end{aligned}
\end{equation}
In the next step, Alice applies the CNOT($A_0$,$a_0$) and CNOT($A_1$,$a_1$) gates, and the system state will become as follows Eq. \eqref{4}. 
\begin{equation}
\begin{aligned}
|\psi'\rangle_{\text{A}_0 \text{A}_1 \text{a}_0 \text{a}_1 \text{b}_0 \text{b}_1} = \frac{1}{2} ( 
\alpha_0 |000000\rangle + \alpha_0 |000101\rangle + \alpha_0 |001010\rangle + \alpha_0 |001111\rangle \\
+ \alpha_1 |101000\rangle + \alpha_1 |101101\rangle + \alpha_1 |100010\rangle + \alpha_1 |100111\rangle \\
+ \alpha_2 |111100\rangle + \alpha_2 |111001\rangle + \alpha_2 |110110\rangle + \alpha_2 |110011\rangle 
)
 \label{4}
\end{aligned}
\end{equation}

\begin{figure}[t]
\centerline{\includegraphics[width=1.000\textwidth]{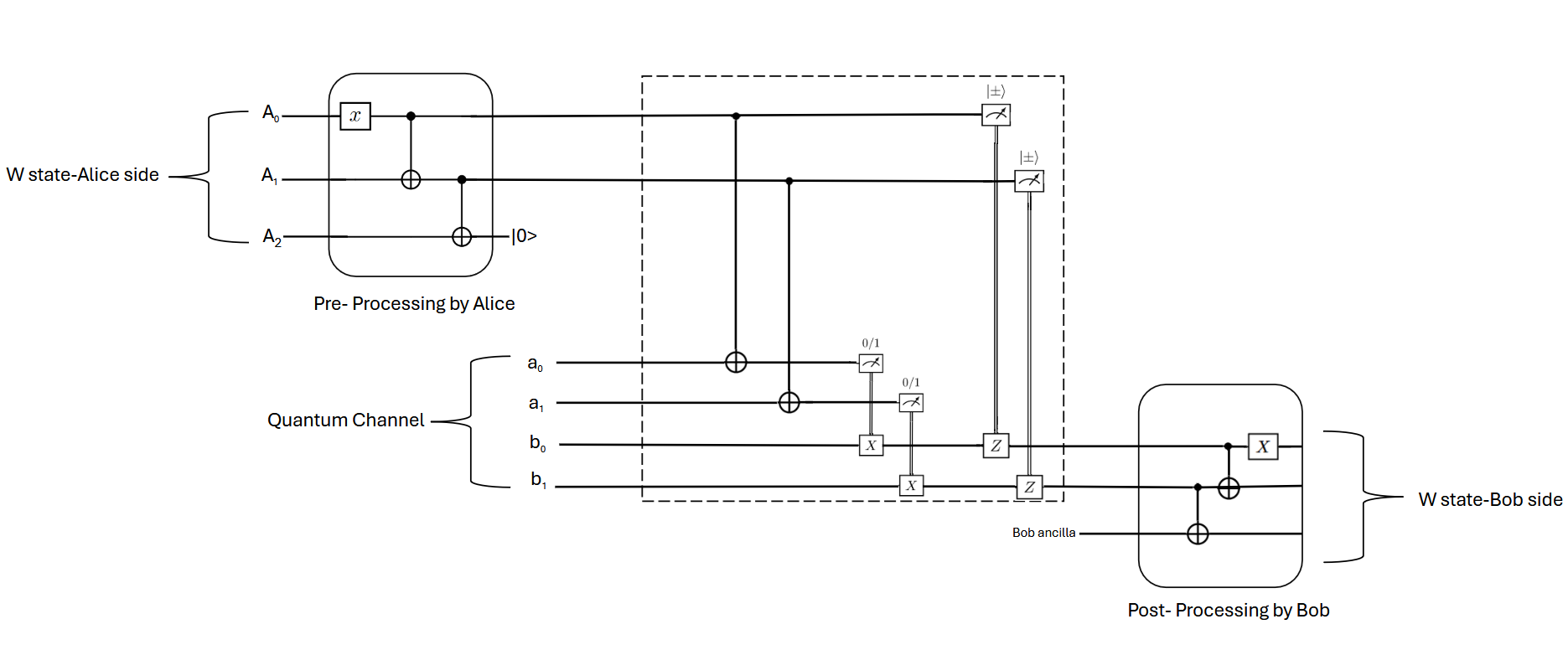}}
\vspace*{8pt}
\caption{ Teleportaion protocol for 3 particle W state}
\label{fig:3}
\end{figure}

\begin{table}[ht]
\centering
\begin{tabular}{c c c}
\hline
\textbf{Alice's results} & \textbf{Bob's states} & \textbf{Bob's operators} \\ 
\textbf{$\text{A}_0 \text{A}_1 \text{a}_0 \text{a}_1$} & \textbf{$  \text{b}_0 \text{b}_1$}  \\
\hline
+ + 00 & \(\alpha_0 |00\rangle + \alpha_1|10\rangle + \alpha_2|11\rangle\) & \((\sigma_I \otimes \sigma_I)(\sigma_I \otimes \sigma_I)\) \\
+ + 01 & \(\alpha_0 |01\rangle + \alpha_1|11\rangle + \alpha_2|10\rangle\) & \((\sigma_I \otimes \sigma_I)(\sigma_I \otimes \sigma_x)\) \\
+ + 10 & \(\alpha_0 |10\rangle + \alpha_1|00\rangle + \alpha_2|01\rangle\) & \((\sigma_I \otimes \sigma_I)(\sigma_x \otimes \sigma_I)\) \\
+ + 11 & \(\alpha_0 |11\rangle + \alpha_1|01\rangle + \alpha_2|00\rangle\) & \((\sigma_I \otimes \sigma_I)(\sigma_x \otimes \sigma_x)\) \\
+ - 00 & \(\alpha_0 |00\rangle + \alpha_1|10\rangle - \alpha_2|11\rangle\) & \((\sigma_I \otimes \sigma_z)(\sigma_I \otimes \sigma_I)\) \\
+ - 01 & \(\alpha_0 |01\rangle + \alpha_1|11\rangle - \alpha_2|10\rangle\) & \((\sigma_I \otimes \sigma_z)(\sigma_I \otimes \sigma_x)\) \\
+ - 10 & \(\alpha_0 |10\rangle + \alpha_1|00\rangle - \alpha_2|01\rangle\) & \((\sigma_I \otimes \sigma_z)(\sigma_x \otimes \sigma_I)\) \\
+ - 11 & \(\alpha_0 |11\rangle + \alpha_1|01\rangle - \alpha_2|00\rangle\) & \((\sigma_I \otimes \sigma_z)(\sigma_x \otimes \sigma_x)\) \\
- + 00 & \(\alpha_0 |00\rangle - \alpha_1|10\rangle - \alpha_2|11\rangle\) & \((\sigma_z \otimes \sigma_I)(\sigma_I \otimes \sigma_I)\) \\
- + 01 & \(\alpha_0 |01\rangle - \alpha_1|11\rangle - \alpha_2|10\rangle\) & \((\sigma_z \otimes \sigma_I)(\sigma_I \otimes \sigma_x)\) \\
- + 10 & \(\alpha_0 |10\rangle - \alpha_1|00\rangle - \alpha_2|01\rangle\) & \((\sigma_z \otimes \sigma_I)(\sigma_x \otimes \sigma_I)\) \\
- + 11 & \(\alpha_0 |11\rangle - \alpha_1|01\rangle - \alpha_2|00\rangle\) & \((\sigma_z \otimes \sigma_I)(\sigma_x \otimes \sigma_x)\) \\
- - 00 & \(\alpha_0 |00\rangle - \alpha_1|10\rangle + \alpha_2|11\rangle\) & \((\sigma_z \otimes \sigma_z)(\sigma_I \otimes \sigma_I)\) \\
- - 01 & \(\alpha_0 |01\rangle - \alpha_1|11\rangle + \alpha_2|10\rangle\) & \((\sigma_z \otimes \sigma_z)(\sigma_I \otimes \sigma_x)\) \\
- - 10 & \(\alpha_0 |10\rangle - \alpha_1|00\rangle + \alpha_2|01\rangle\) & \((\sigma_z \otimes \sigma_z)(\sigma_x \otimes \sigma_I)\) \\
- - 11 & \(\alpha_0 |11\rangle - \alpha_1|01\rangle + \alpha_2|00\rangle\) & \((\sigma_z \otimes \sigma_z)(\sigma_x \otimes \sigma_x)\) \\
\hline
\end{tabular}
\caption{Bob's states and operators for the system are determined based on Alice's measurements}
\end{table}
After applying the operators in Table 1, Bob's quantum state becomes $|\psi''\rangle_{b_0 b_1}  = \alpha_0 |00\rangle + \alpha_1 |10\rangle + \alpha_2 |11\rangle$.Then, using an ancilla qubit, Bob, through post-processing on his qubits, which includes the  X(first qubit), CNOT(first qubit, second qubit), and CNOT (second qubit, ancilla qubit) gates, reconstructs Alice's initial quantum state on his own qubits.
$|\psi'''\rangle_{b_0 b_1 b_{ancilla}}  = \alpha_0 |100\rangle + \alpha_1 |010\rangle + \alpha_2 |001\rangle$ \\
The schematic of this protocol is shown in Fig.\ref{fig:3}.

\section{Teleportation   of a Four-Particle W State}
Consider the  unkhown quantum  Four-Particle W State  $|\phi\rangle_{\text{B}_0   \text{B}_1 \text{B}_2 \text{B}_3 } = \beta_0 |0010\rangle + \beta_1 |0100\rangle + \beta_2 |1000\rangle + \beta_3 |0001\rangle $ on Bob's qubits.Initially, a pre-processing step is performed by Bob on the unknown qubits, including the CNOT(fourth qubit , second qubit),CNOT(fourth qubit , firest qubit).TOFFOLI(firest qubit , secind qubit , fourth qubit),X(firest qubit),X(second qubit),TOFFOLI(firest qubit , second qubit , third qubit),X(firest qubit),X(second qubit) gates. After this pre-processing, two of the unknown qubits are reset to $|0\rangle$. Now, Bob only needs to send the remaining two qubits through the $|\omega\rangle_{\text{b}_0 \text{b}_1 \text{a}_0 \text{a}_1} $  channel , this channel is no different from the channel in the previous section.
The general state of the system can be expressed according to Eq. \eqref{5}.
\begin{equation}
\begin{aligned}
|\psi\rangle_{\text{B}_0 \text{B}_1 \text{b}_0 \text{b}_1 \text{a}_0 \text{a}_1} = \frac{1}{2} \big( 
& \beta_0 |000000\rangle + \beta_0 |000101\rangle + \beta_0 |001010\rangle + \beta_0 |001111\rangle \\
& + \beta_1 |010000\rangle + \beta_1 |010101\rangle + \beta_1 |011010\rangle + \beta_1 |011111\rangle \\
& + \beta_2 |100000\rangle + \beta_2 |100101\rangle + \beta_2 |101010\rangle + \beta_2 |101111\rangle \\
& + \beta_3 |110000\rangle + \beta_3 |110101\rangle + \beta_3 |111010\rangle + \beta_3 |111111\rangle 
\big)
\label{5}
\end{aligned}
\end{equation}
In the next step, Bob applies the CNOT($B_0$,$b_0$) and CNOT($B_1$,$b_1$) gates, and the system state will become as follows Eq. \eqref{6}. 
\begin{equation}
\begin{aligned}
|\psi'\rangle_{\text{B}_0 \text{B}_1 \text{b}_0 \text{b}_1 \text{a}_0 \text{a}_1} = \frac{1}{2} \big( 
& \beta_0 |000000\rangle + \beta_0 |000101\rangle + \beta_0 |001010\rangle + \beta_0 |001111\rangle \\
& + \beta_1 |010100\rangle + \beta_1 |010001\rangle + \beta_1 |011110\rangle + \beta_1 |011011\rangle \\
& + \beta_2 |101000\rangle + \beta_2 |101101\rangle + \beta_2 |100010\rangle + \beta_2 |100111\rangle \\
& + \beta_3 |111100\rangle + \beta_3 |111001\rangle + \beta_3 |110110\rangle + \beta_3 |110011\rangle 
\big)
\label{6}
\end{aligned}
\end{equation}

then, Bob performs X basis measurements (Hadamard basis) on her qubits $B_0$ and $B_1$, and Z basis measurements (standard basis) on the qubits of her channel, Table 2 shows the system's collapsed states and the unitary operators that Alice must apply to reconstruct her qubits.

\begin{table}[ht]
\centering
\begin{tabular}{c c c}
\hline
\textbf{Bob's results} & \textbf{Aice's states} & \textbf{Alice's operators} \\ 
\textbf{$\text{B}_0 \text{B}_1 \text{b}_0 \text{b}_1$} & \textbf{$\text{a}_0 \text{a}_1$}  \\
\hline
+ + 00 & \( \beta_0 |00\rangle + \beta_1|01\rangle + \beta_2|10\rangle + \beta_3|11\rangle \) & \( (\sigma_I \otimes \sigma_I)(\sigma_I \otimes \sigma_I) \) \\
+ + 01 & \( \beta_0 |01\rangle + \beta_1|00\rangle + \beta_2|11\rangle + \beta_3|10\rangle \) & \( (\sigma_I \otimes \sigma_I)(\sigma_I \otimes \sigma_x) \) \\
+ + 10 & \( \beta_0 |10\rangle + \beta_1|11\rangle + \beta_2|00\rangle + \beta_3|01\rangle \) & \( (\sigma_I \otimes \sigma_I)(\sigma_x \otimes \sigma_I) \) \\
+ + 11 & \( \beta_0 |11\rangle + \beta_1|10\rangle + \beta_2|01\rangle + \beta_3|00\rangle \) & \( (\sigma_I \otimes \sigma_I)(\sigma_x \otimes \sigma_x) \) \\
+ - 00 & \( \beta_0 |00\rangle - \beta_1|01\rangle + \beta_2|10\rangle - \beta_3|11\rangle \) & \( (\sigma_I \otimes \sigma_z)(\sigma_I \otimes \sigma_I) \) \\
+ - 01 & \( \beta_0 |01\rangle - \beta_1|00\rangle + \beta_2|11\rangle - \beta_3|10\rangle \) & \( (\sigma_I \otimes \sigma_z)(\sigma_I \otimes \sigma_x) \) \\
+ - 10 & \( \beta_0 |10\rangle - \beta_1|11\rangle + \beta_2|00\rangle - \beta_3|01\rangle \) & \( (\sigma_I \otimes \sigma_z)(\sigma_x \otimes \sigma_I) \) \\
+ - 11 & \( \beta_0 |11\rangle - \beta_1|10\rangle + \beta_2|01\rangle - \beta_3|00\rangle \) & \( (\sigma_I \otimes \sigma_z)(\sigma_x \otimes \sigma_x) \) \\
- + 00 & \( \beta_0 |00\rangle + \beta_1|01\rangle - \beta_2|10\rangle - \beta_3|11\rangle \) & \( (\sigma_z \otimes \sigma_I)(\sigma_I \otimes \sigma_I) \) \\
- + 01 & \( \beta_0 |01\rangle + \beta_1|00\rangle - \beta_2|11\rangle - \beta_3|10\rangle \) & \( (\sigma_z \otimes \sigma_I)(\sigma_I \otimes \sigma_x) \) \\
- + 10 & \( \beta_0 |10\rangle + \beta_1|11\rangle - \beta_2|00\rangle - \beta_3|01\rangle \) & \( (\sigma_z \otimes \sigma_I)(\sigma_x \otimes \sigma_I) \) \\
- + 11 & \( \beta_0 |11\rangle + \beta_1|10\rangle - \beta_2|01\rangle - \beta_3|00\rangle \) & \( (\sigma_z \otimes \sigma_I)(\sigma_x \otimes \sigma_x) \) \\
- - 00 & \( \beta_0 |00\rangle - \beta_1|01\rangle - \beta_2|10\rangle + \beta_3|11\rangle \) & \( (\sigma_z \otimes \sigma_z)(\sigma_I \otimes \sigma_I) \) \\
- - 01 & \( \beta_0 |01\rangle - \beta_1|00\rangle - \beta_2|11\rangle + \beta_3|10\rangle \) & \( (\sigma_z \otimes \sigma_z)(\sigma_I \otimes \sigma_x) \) \\
- - 10 & \( \beta_0 |10\rangle - \beta_1|11\rangle - \beta_2|00\rangle + \beta_3|01\rangle \) & \( (\sigma_z \otimes \sigma_z)(\sigma_x \otimes \sigma_I) \) \\
- - 11 & \( \beta_0 |11\rangle - \beta_1|10\rangle - \beta_2|01\rangle + \beta_3|00\rangle \) & \( (\sigma_z \otimes \sigma_z)(\sigma_x \otimes \sigma_x) \) \\
\hline
\end{tabular}
\caption{Bob's states and operators for the system are determined based on Alice's measurements}
\end{table}

\begin{figure}[t]
\centerline{\includegraphics[width=1.100\textwidth]{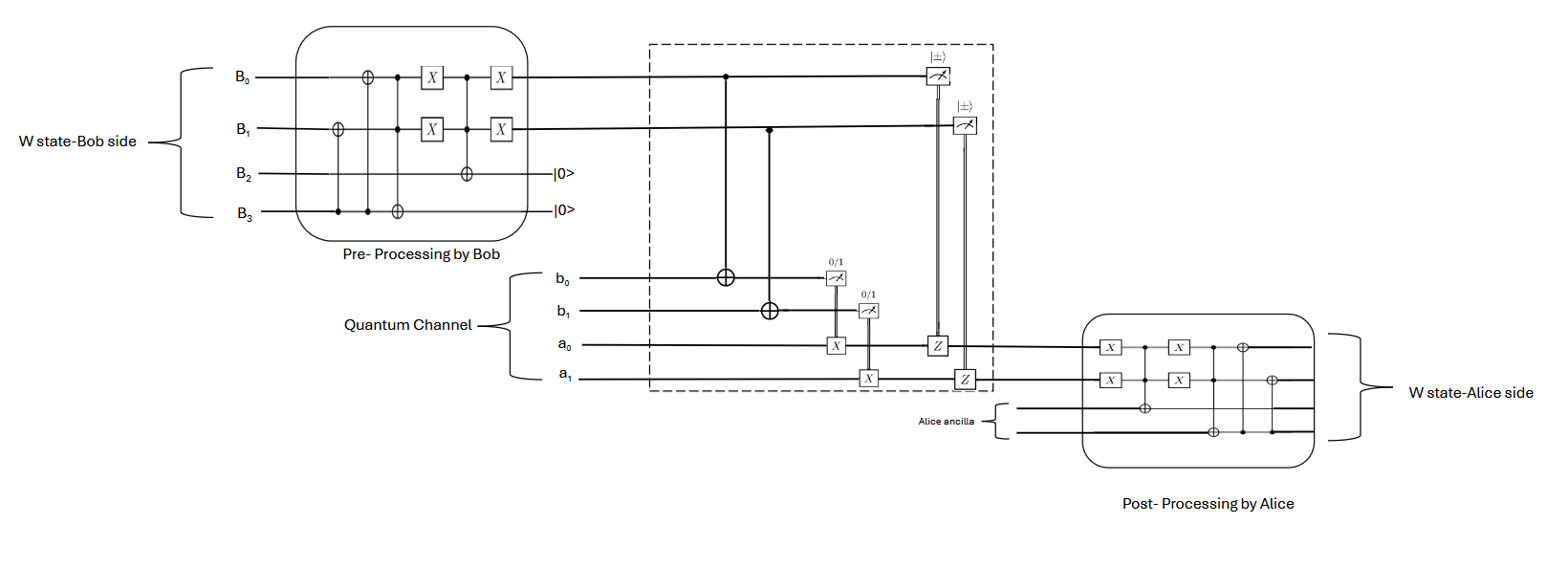}}
\vspace*{8pt}
\caption{ Teleportaion protocol for 4 particle W state}
\label{fig:4}
\end{figure}

After reconstructing the qubits,Alice performs post-processing on his qubits using two ancilla qubits to construct the final state, as shown in Fig.\ref{fig:4}.

\section{Bidirectional  teleportation of three and four particle  W state}
\begin{figure}[t]
\centerline{\includegraphics[width=1.100\textwidth]{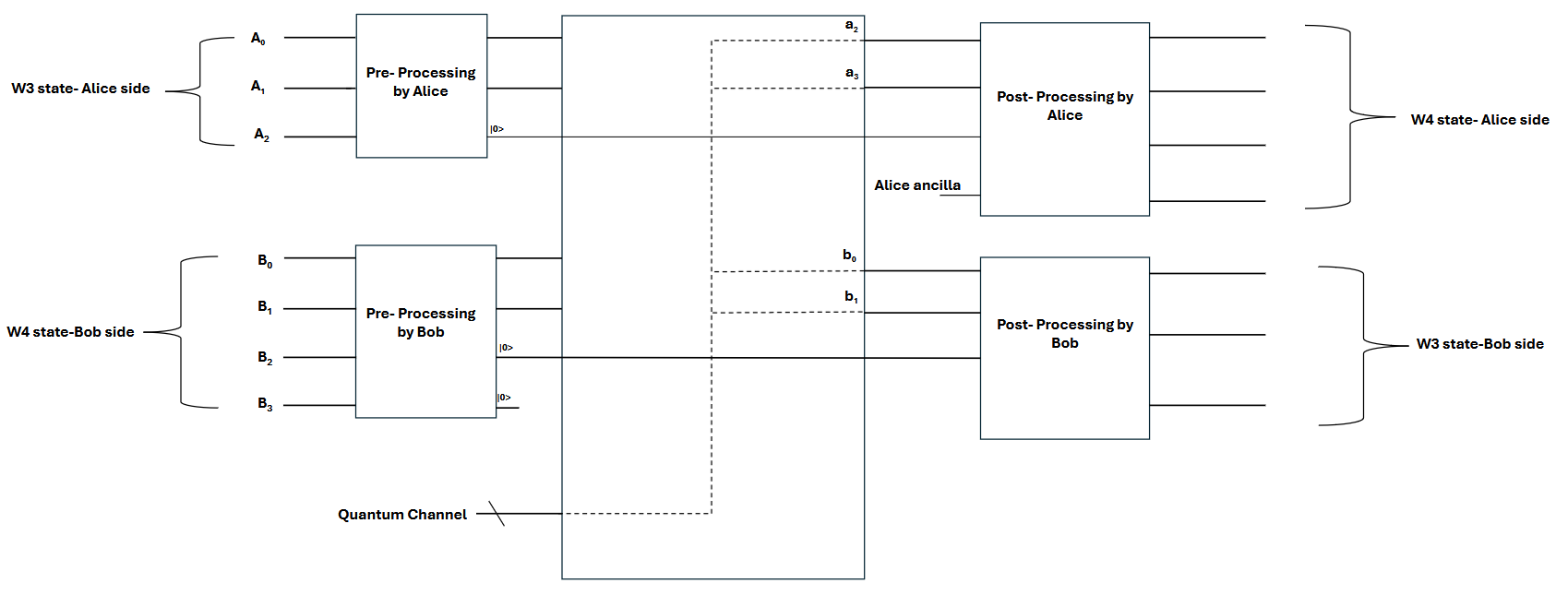}}
\vspace*{8pt}
\caption{schematic of bidirectional teleportation}
\label{fig:5}
\end{figure}

Alice wants to send the $|\psi\rangle_{\text{A}_0 \text{A}_1 \text{A}_3}= \alpha_0 |100\rangle + \alpha_1 |010\rangle + \alpha_2 |001\rangle$  state  to bob . After performing the pre-processing introduced in Section four, her state transforms into $|\psi\rangle_{\text{A}_0 \text{A}_1  \text{A}_3 }=( \alpha_0 |00\rangle + \alpha_1 |10\rangle + \alpha_2 |11\rangle)|0\rangle $
.On the other side, Bob has the $|\phi\rangle_{\text{B}_0   \text{B}_1 \text{B}_2 \text{B}_3 } = \beta_0 |0010\rangle + \beta_1 |0100\rangle + \beta_2 |1000\rangle + \beta_3 |0001\rangle $
state and simultaneously applies the preprocessing introduced in Section five to his qubits to transform into the $|\phi\rangle_{\text{B}_0   \text{B}_1 \text{B}_2 \text{B}_3 } = (\beta_0 |00\rangle + \beta_1 |01\rangle + \beta_2 |10\rangle + \beta_3 |11\rangle) |00\rangle$ \\
For the quantum channel, Alice and Bob each share four qubits, the $ a_0 a_1 a_2 a_3$ belong to Alice ,while the $ b_0 b_1 b_2 b_3$ qubits belong to Bob, And sequentially, the CNOT gates is applied between Alice's qubits as the control and Bob's qubits as the target, so that the channel state can be expressed as $|\omega\rangle_{\text{a}_0   \text{a}_1 \text{a}_2 \text{a}_3  \text{b}_0   \text{b}_1 \text{b}_2 \text{b}_3} =\frac{1}{4} \sum_{i=0}^{1} \sum_{j=0}^{1} \sum_{k=0}^{1} \sum_{l=0}^{1} |i j k l\rangle |i j k l\rangle $ \\
 In the next step, Alice applies the  $CNOT(A_0 , a_0)$ , $CNOT(A_1,a_1)$ And Bob applies the  $CNOT(B_0 , b_2)$ , $CNOT(B_1,b_3)$And they also perform measurements in the $X$ basis on $A_0 A_1 B_0 B_1$ And  measure the $a_0 a_1 b_2 b_3$ qubits in the $Z$ basis.
Then, based on Bob's results and regardless of her own measurement outcomes, Alice can refer to Table 2 And apply the unitary operators on the $a_2 a_3$ qubits,Bob also simultaneously refers to Table 1 and, based on Alice's results, applies the unitary operators to the $b_0 b_1$ .After applying the unitary gates, Alice uses the $A_2$ qubit and an ancilla qubit, while Bob uses the $B_2$ qubit to reconstruct the qubits through post-processing.\\
The schematic of the above protocol is shown in Fig.\ref{fig:5}.

\section{Conclusion}
Based on what is stated in article \cite{Yuan2008} , equation $\eta = \frac{q_s}{q_u + b_t + q_a} $is used to compare the efficiency between  several  protocols. In this equation, $q_s$ represents the number of qubits transmitted, $q_u$ denotes quantum resources,$ b_t $represents classical resources, and $q_a $corresponds to ancilla qubits.
In Table 3, the efficiency of articles \citen{Nie2010,Mandal2023,Sang2015,sadeghi2017} has been calculated and compared with the efficiency of the protocols presented in this paper, demonstrating that our proposed protocols achieve higher efficiency.

\begin{table}[ht]

\scriptsize
\centering
\begin{tabular}{c c c c c}
\hline
\textbf{Protocol} & \textbf{Transmitted Qubits} & \textbf{Quantum Channel}&\textbf{Classical Bits}&\textbf{Efficiency}\\
\hline
\\

\cite{Nie2010}  & three-qubit state&two four-qubit cluster states&8& $\frac{3}{16}\simeq18.7\%$ \\
\\

\cite{Mandal2023} & three-particle W state &seven-qubit cluster state&7& $\frac{3}{14}\simeq21.4\%$\\
\\
\cite{Sang2015} &one-two GHZ state (bidirectional) &five-qubit Cluster State&5& $\frac{3}{10}=30\%$\\
\\
\cite{sadeghi2017}&two-two qubit state (bidirectional)&eight-Qubit Entangled State &8& $\frac{4}{16}=25\%$\\
\\

This paper&three-particle W state&four qubit cluster state+ 1 ancilla qubit&4& $\frac{3}{9}\simeq33.3\%$ \\
\\

This paper&four- particle W state&four qubit cluster state+ 2 ancilla qubits&4& $\frac{4}{10}=40\%$ \\
\\
This paper&three-four particle W state(bidirectional)&eight qubit cluster state+ 1 ancilla qubit&4&$\frac{7}{17}\simeq41.1\%$ \\
\hline
\end{tabular}
\caption{Comparison of different protocols with our protocols in efficiency}
\end{table}

\end{document}